\documentclass[aps,pra,twocolumn,showpacs,amsmath,amssymb,superscriptaddress]{revtex4-1}

\usepackage{graphicx}
\usepackage{dcolumn}
\usepackage{bm}

\begin{document}

\title{Dissipative phase transitions: Independent versus collective decay and spin squeezing}

\author{Tony E. Lee}
\affiliation{ITAMP, Harvard-Smithsonian Center for Astrophysics, Cambridge, Massachusetts 02138, USA}
\affiliation{Department of Physics, Harvard University, Cambridge, Massachusetts 02138, USA}
\author{Ching-Kit Chan}
\affiliation{ITAMP, Harvard-Smithsonian Center for Astrophysics, Cambridge, Massachusetts 02138, USA}
\affiliation{Department of Physics, Harvard University, Cambridge, Massachusetts 02138, USA}
\author{Susanne F. Yelin}
\affiliation{ITAMP, Harvard-Smithsonian Center for Astrophysics, Cambridge, Massachusetts 02138, USA}
\affiliation{Department of Physics, Harvard University, Cambridge, Massachusetts 02138, USA}
\affiliation{Department of Physics, University of Connecticut, Storrs, Connecticut 06269, USA}

\date{\today}

\begin{abstract}
We study the XY model with infinite-range interactions (Lipkin-Meshkov-Glick model) in the presence of dissipation from spontaneous decay. We show that independent and collective decay lead to qualitatively different phase transitions of the steady state, even though the phase boundary is the same. Independent decay leads to a second-order phase transition to a ferromagnet, while collective decay leads to a first-order transition to a time-dependent oscillatory phase. Then we show that the addition of a drive leads to infinite spin squeezing for collective decay in the thermodynamic limit. Our results can be experimentally seen in trapped-ion and cavity-QED experiments.
\end{abstract}

\maketitle

\section{Introduction}

In recent years, there has been interest in phase transitions of atomic ensembles in the presence of dissipation from spontaneous decay \cite{daley14,armen06,lee11,lee12,schonleber14,vidanovic14,jin14,lee13c,joshi13,  morrison08a,morrison08b,bohnet12,kessler12,dallatorre13,gonzalez13,wolfe14b,ritsch13,carr13a,malossi14,olmos14,lee14a,xu14,lang14}. A motivation is that less is known about nonequilibrium systems than equilibrium ones, so it is of fundamental interest to see what new behavior arises due to dissipation. Another motivation is that dissipative systems can exhibit a large amount of spin squeezing \cite{kessler12,dallatorre13,gonzalez13,wolfe14b} and may thus be useful for quantum metrology \cite{wineland92,kitagawa93,kuzmich00,takano09,schleiersmith10b,schleiersmith10a,leroux10,wasilewski10,sewell12}.



Spontaneous decay in an ensemble of atoms can be either independent or collective. If the atoms are in free space with a spacing much larger than a wavelength, the decay is independent for each atom. If the atoms are coupled to a lossy cavity \cite{bonifacio71,bohnet12} or close to each other \cite{gross82,lin12}, the decay is collective (superradiance). There has been work on phase transitions due to either independent decay \cite{armen06,lee11,lee12,schonleber14,vidanovic14,jin14,lee13c,joshi13} or collective decay \cite{morrison08a,morrison08b,bohnet12,kessler12,dallatorre13,gonzalez13,wolfe14b,ritsch13,carr13a,malossi14,olmos14,lee14a,xu14,lang14}. This raises the question of how the phase transitions of each type are related to each other.

\begin{figure}[t]
\centering
\includegraphics[width=3.5 in,trim=1.in 4.1in 1in 4.1in,clip]{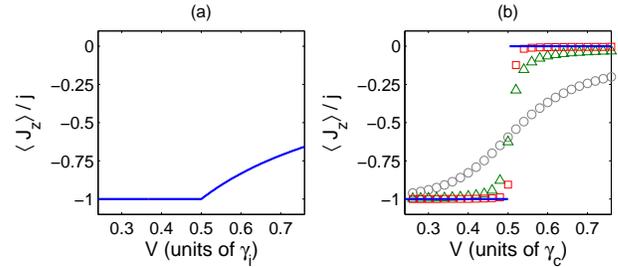}
\caption{\label{fig:jz}$\langle J_z\rangle/j$ with $j=N/2$ for (a) independent decay and (b) collective decay. Solid lines are mean-field predictions. Numerical results are shown for different numbers of atoms: $N=10$ (gray circles), $N=100$ (green triangles), and $N=1000$ (red squares). The phase transition is at $|V|=\gamma/2$ for both types of decay.}
\end{figure}

In this paper, we make a direct comparison between independent and collective decay for the $XY$ model with infinite-range interactions (Lipkin-Meshkov-Glick model \cite{lipkin65}). Independent decay was previously shown to cause a second-order phase transition from a paramagnet to a ferromagnet [Fig.~\ref{fig:jz}(a)] \cite{lee13c,joshi13}. Here, we show that collective decay qualitatively changes the phase transition. Although the phase boundary is the same, there is now a first-order transition to a time-dependent oscillatory phase [Fig.~\ref{fig:jz}(b)]. The differences are due to the fact that collective decay conserves angular momentum, causing the Bloch vector to have maximal length. However, both types of decay have spin squeezing limited to $\xi^2\geq 1/2$, where $\xi^2$ is the spin-squeezing parameter \cite{wineland92}.

Then we show that the addition of a drive leads to maximal spin squeezing ($\xi^2\rightarrow 0$) for collective decay in the thermodynamic limit. We also consider limitations due to finite-size effects and independent decay. Our work complements previous work on dissipative spin squeezing \cite{kessler12,dallatorre13,gonzalez13,wolfe14b}. We note that the squeezing here is not due to a dark state as in Ref.~\cite{dallatorre13}.

Our results can be experimentally seen using trapped ions or cavity QED. With trapped ions, the spin-spin interactions are mediated by motional modes \cite{molmer99,richerme14}, and collective decay is mediated by an auxiliary ion \cite{schneider02,lin13}. With atoms in a cavity, the cavity mediates the spin-spin interaction as well as the collective decay \cite{morrison08a,morrison08b}.

The paper is outlined as follows. In Sec.~\ref{sec:model}, we define the Hamiltonian and the master equations that we study. In Sec.~\ref{sec:indep}, we review results for independent decay. In Sec.~\ref{sec:coll}, we present results for collective decay. In Sec.~\ref{sec:drive} we discuss what happens in the presence of a drive. In Appendix \ref{sec:app}, we provide details on the spin-squeezing calculations.

\section{Model} \label{sec:model}

We consider the anisotropic-$XY$ model with infinite-range interactions,
\begin{eqnarray}
H&=&\frac{V}{N}(J_x^2 - J_y^2), \label{eq:H_xy} \\
&=&\frac{V}{2N}(J_+^2 + J_-^2), \label{eq:H_xy_JpJm}
\end{eqnarray}
where $\vec{J}=\frac{1}{2}\sum_n \vec{\sigma}^n$ and $J_\pm=\sum_n \sigma^n_\pm$ are collective spin operators, and $N$ is the number of atoms. As seen from Eq.~\eqref{eq:H_xy_JpJm}, the anisotropic interaction excites two spins at a time.

Equation \eqref{eq:H_xy} is a special case of the Lipkin-Meshkov-Glick model \cite{lipkin65}, since we assume maximum anisotropy between $J_x^2$ and $J_y^2$ terms. The equilibrium ground state has been studied in terms of phase transitions \cite{botet83} and spin-squeezing \cite{vidal06,ma09}. Equation \eqref{eq:H_xy} is also known as the two-axis countertwisting model and leads to spin squeezing during the time evolution \cite{kitagawa93}. The model can be experimentally implemented using trapped ions \cite{molmer99,richerme14} or atoms in a cavity \cite{morrison08a,morrison08b}: the ion motion or the cavity mediates an interaction that excites two spins at a time.

We are interested in the nonequilibrium behavior that arises due to dissipation from spontaneous decay of the atoms. In this case, the system is described by a master equation for the density matrix $\rho$. We first consider independent decay,
\begin{eqnarray}
\dot\rho&=&-i[H,\rho]+\frac{\gamma_i}{2}\sum_n(2\sigma^n_-\rho\sigma^n_+ - \sigma^n_+\sigma^n_-\rho - \rho \sigma^n_+\sigma^n_-), \quad \label{eq:master_indep}
\end{eqnarray}
where $\sigma^n_\pm=(\sigma^n_x \pm i\sigma^n_y)/2$. Then we consider collective decay (Dicke superradiance \cite{gross82}),
\begin{eqnarray}
\dot\rho&=&-i[H,\rho]+\frac{\gamma_c}{2N}(2J_-\rho J_+ - J_+J_-\rho - \rho J_+J_-). \label{eq:master_coll}
\end{eqnarray}
$\gamma_i$ and $\gamma_c$ are the rates of independent and collective decay, respectively.

In the trapped-ion implementation, the decay can be either independent or collective: independent decay is from optical pumping, while collective decay is mediated by an auxiliary ion \cite{schneider02,lin13}. In the cavity implementation, the cavity mediates a collective decay \cite{morrison08a,morrison08b}, although there is an ``upward'' decay in addition to the downward decay in Eq.~\eqref{eq:master_coll} \footnote{When there is both upward and downward decay, the mean-field equations are the same as Eqs.~\eqref{eq:coll_x}--\eqref{eq:coll_z} but with $\gamma_c$ replaced by the net downward decay rate.}. Note that even if collective decay dominates ($\gamma_c\gg\gamma_i$), independent decay events will eventually occur. In this paper, when we consider
collective decay, we are referring to time scales shorter than $1/\gamma_i$. We discuss the modification by independent decay in Sec.~\ref{sec:limitations}.

The steady states of the master equations exhibit phase transitions as the parameters are varied. It is important to note that both master equations have a $Z_2$ symmetry: $\sigma^n_x,\sigma^n_y\rightarrow-\sigma^n_x,-\sigma^n_y$. This is the symmetry that is broken in the ordered phase.


The intuition for the phase transition is as follows. Since the anisotropic interaction excites two spins at a time [Eq.~\eqref{eq:H_xy_JpJm}], there is competition between the pairwise excitation and the decay. If the interaction is weak, the decay dominates, and the Bloch vector points downwards in steady state. But if the interaction is strong enough, the pairwise excitation dominates, and the Bloch vector points sideways. Thus, there is a phase transition when $|V|$ is sufficiently large.


Throughout the paper, we characterize the steady states in terms of their spin squeezing. We use the spin-squeezing parameter as defined by Wineland \emph{et al.}~\cite{wineland92},
\begin{eqnarray}
\xi^2&=&\min_{\vec{n}_\perp}\, \frac{N(\Delta J_{\vec{n}_\perp})^2}{|\langle \vec{J} \rangle|^2}, \label{eq:xi2_wineland}
\end{eqnarray}
where we minimize with respect to unit vectors $\vec{n}_\perp$ that are normal to $\langle \vec{J} \rangle$. When $\xi^2<1$, the spins have improved phase sensitivity to rotations compared to the shot-noise limit, and are thus useful for metrology. Note that if the Bloch vector has maximal length $|\langle \vec{J} \rangle|=N/2$, Eq.~\eqref{eq:xi2_wineland} coincides with the definition by Kitagawa and Ueda \cite{kitagawa93}.

We note that Ref.~\cite{baragiola10} includes a comparison of spin squeezing in the presence of collective or independent decay. Here, we systematically compare the two cases by analytically calculating $\xi^2$.

\section{Independent decay} \label{sec:indep}

The case of independent decay was previously considered \cite{lee13c,joshi13}, and we summarize it here. To obtain the mean-field equations, we find the equations of motion for $\vec{J}$ and then factorize terms like $\langle J_xJ_y\rangle=\langle J_x\rangle\langle J_y\rangle$. For convenience, we define $X=\langle J_x\rangle/j$, $Y=\langle J_y\rangle/j$, $Z=\langle J_z\rangle/j$ where $j=N/2$, so that $X,Y,Z\in[-1,1]$. The mean-field equations for Eq.~\eqref{eq:master_indep} are
\begin{eqnarray}
\dot{X}&=&-V YZ -\frac{\gamma_i}{2}X,\label{eq:indep_x}\\
\dot{Y}&=&-V XZ - \frac{\gamma_i}{2}Y,\\
\dot{Z}&=&2VXY - \gamma_i(Z+1).\label{eq:indep_z}
\end{eqnarray}

To find the phases, we solve for the fixed points of Eqs.~\eqref{eq:indep_x}--\eqref{eq:indep_z}. When $|V|<\gamma_i/2$, the steady state is
\begin{eqnarray}
\bar{X}=\bar{Y}=0, \quad \bar{Z}=-1,
\end{eqnarray}
which means that the Bloch vector points downwards. We call this phase the paramagnet (PM) since it does not break the $Z_2$ symmetry. When $|V|>\gamma_i/2$, there are two possible steady states:
\begin{eqnarray}
\bar{X}&=&\pm\frac{\sqrt{\gamma_i(2|V|-\gamma_i)}}{2V}, \label{eq:xss_indep}\\
\bar{Y}&=&\pm\text{sgn}(V)\frac{\sqrt{\gamma_i(2|V|-\gamma_i)}}{2V}, \label{eq:yss_indep}\\
\bar{Z}&=&-\frac{\gamma_i}{2|V|}, \label{eq:zss_indep}
\end{eqnarray}
so the Bloch vector points sideways. This phase breaks the symmetry, so we call it the ferromagnet (FM). The transition from PM to FM at $|V|=\gamma_i/2$ is second order, since $\bar{X},\bar{Y},\bar{Z}$ are continuous there [Fig.~\ref{fig:jz}(a)]. Note that both phases are time-independent, i.e., stable fixed points.

In the limit of large $N$, one can calculate the spin squeezing analytically by considering fluctuations around the mean-field steady states. In the PM, the spin-squeezing parameter is \cite{lee13c}
\begin{eqnarray}
\xi^2&=&\frac{\gamma_i}{\gamma_i+2|V|}, \quad\quad |V|<\frac{\gamma_i}{2}. \label{eq:xi_indep}
\end{eqnarray}
The squeezing is maximum ($\xi^2$ is minimum) at the phase transition, where $\xi^2=1/2$.




\section{Collective decay} \label{sec:coll}

Now we consider what happens with collective decay. We note that phase transitions of a related model were studied in Refs.~\cite{morrison08a,morrison08b}, but the model we study turns out to have a very different ordered phase.

We assume that the atoms are in the Dicke manifold with maximum angular momentum ($j=N/2$) \cite{gross82}. From Eq.~\eqref{eq:master_coll}, the equations of motion for the spin operators are
\begin{eqnarray}
\partial_t \langle J_x \rangle&=&-\frac{V}{N}\langle \{J_y,J_z\} \rangle + \frac{\gamma_c}{2N} (\langle \{J_x,J_z\} \rangle - \langle J_x \rangle) ,\quad\\
\partial_t \langle J_y \rangle&=&-\frac{V}{N}\langle \{J_x,J_z\} \rangle + \frac{\gamma_c}{2N} (\langle \{J_y,J_z\} \rangle - \langle J_y \rangle) ,\quad\\
\partial_t \langle J_z \rangle&=&\frac{2V}{N}\langle \{J_x,J_y\} \rangle - \frac{\gamma_c}{N}  (\langle J_x^2 \rangle + \langle J_y^2 \rangle + \langle J_z \rangle) .\quad
\end{eqnarray}
Since these equations do not close, it is useful to make a mean-field approximation (factorize terms like $\langle J_xJ_y\rangle=\langle J_x\rangle\langle J_y\rangle$ and take the limit $N\rightarrow\infty$), which is accurate for large $N$. In Sec.~\ref{sec:master}, we compare the mean-field predictions with the original quantum model.

\subsection{Mean-field theory}

The mean-field equations for Eq.~\eqref{eq:master_coll} are
\begin{eqnarray}
\dot{X}&=&-V YZ +\frac{\gamma_c}{2}XZ, \label{eq:coll_x}\\
\dot{Y}&=&-V XZ + \frac{\gamma_c}{2}YZ, \label{eq:coll_y}\\
\dot{Z}&=&2VXY - \frac{\gamma_c}{2}(1-Z^2). \label{eq:coll_z}
\end{eqnarray}
Thus, collective decay leads to nonlinear dissipative terms in the mean-field equations \footnote{The decay in Eq.~\eqref{eq:coll_z} is maximum when $Z=0$ due to collective enhancement \cite{gross82}}, while independent decay leads to linear dissipative terms. Since the master equation conserves angular momentum $\vec{J}^{\,2}$, there is the additional constraint, $X^2+Y^2+Z^2=1$. (This constraint is a key difference with independent decay, as shown below.)



\begin{figure}[tb]
\centering
\begin{tabular}{c}
\includegraphics[width=3.5 in,trim=1.2in 3.4in 1.2in 3.5in,clip]{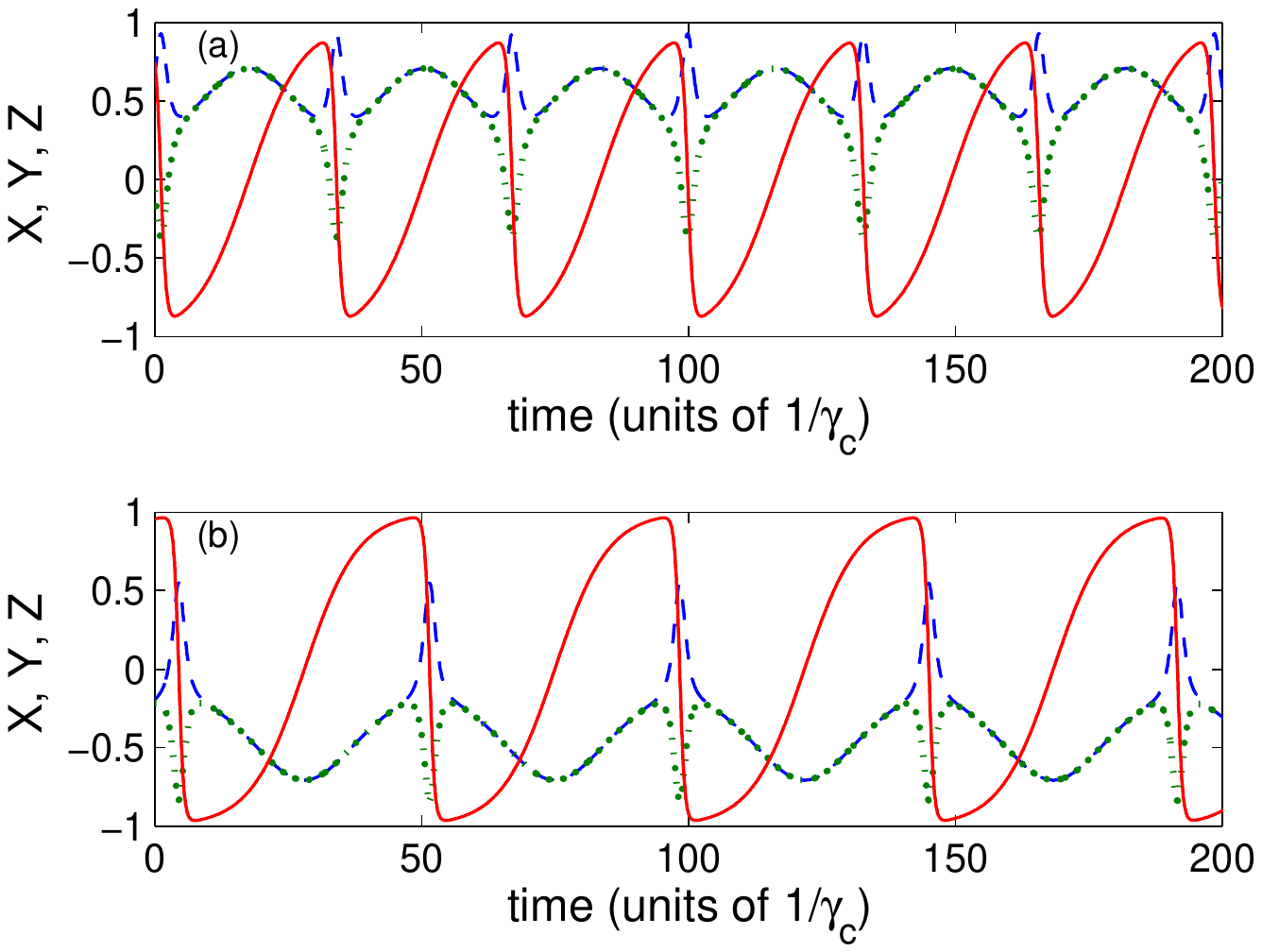}\\
\includegraphics[width=3.5 in,trim=1.2in 4.1in 1.2in 4.2in,clip]{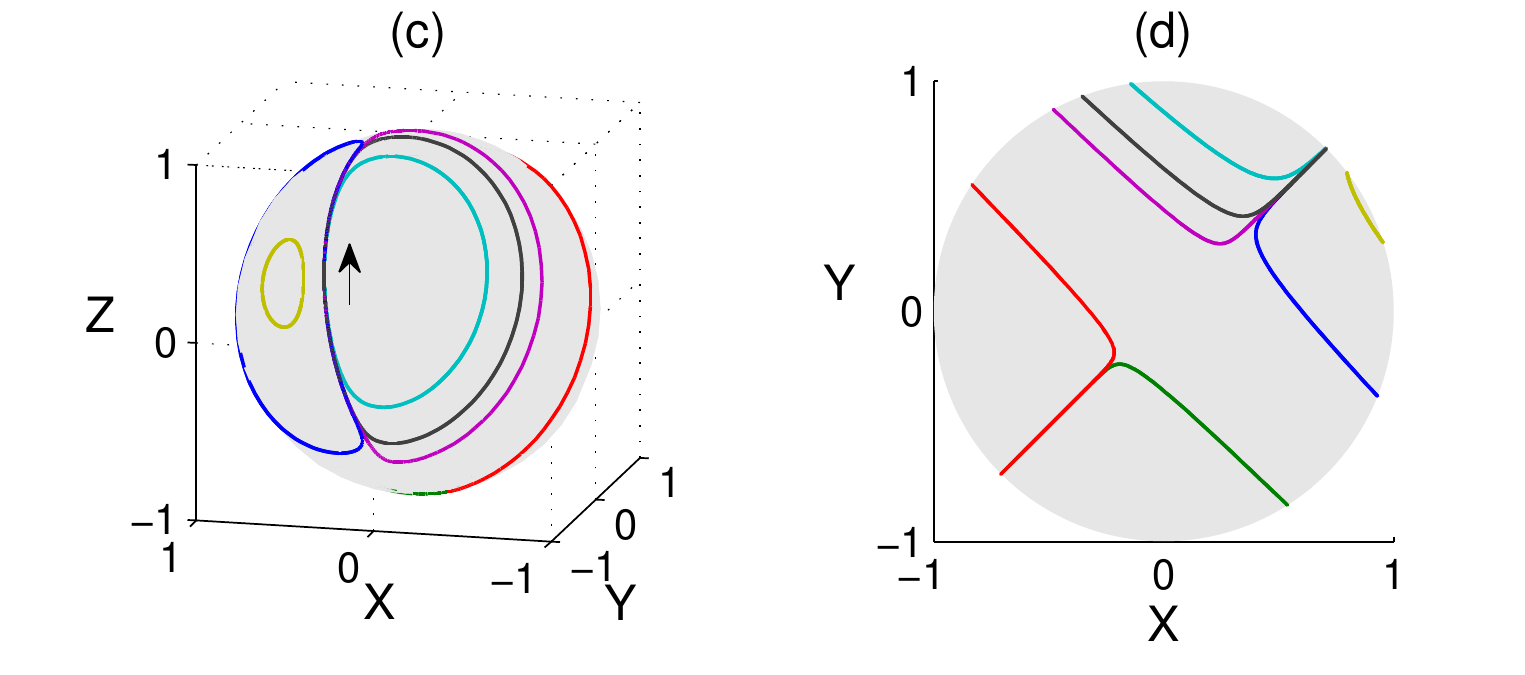}
\end{tabular}
\caption{\label{fig:trajectories}Examples of mean-field periodic orbits for $V=0.6\gamma_c$. (a,b) Trajectories in time showing $X$ (blue dashed line), $Y$ (green dotted line), and $Z$ (red solid line). Panels (a) and (b) are for different initial conditions. (c,d) Trajectories on the Bloch sphere and projected onto the $xy$ plane. Different colors correspond to different initial conditions. The black arrow shows the direction of the trajectories.}
\end{figure}

We solve for the steady states of Eqs.~\eqref{eq:coll_x}--\eqref{eq:coll_z}. When $|V|<\gamma_c/2$, the steady state is
\begin{eqnarray}
\bar{X}=\bar{Y}=0, \quad \bar{Z}=-1,
\end{eqnarray}
which we again call the PM phase. When $|V|>\gamma_c/2$, it turns out that there are no stable fixed points, but there are four center fixed points \cite{strogatz01}. the steady states are periodic orbits, meaning that $X,Y,Z$ oscillate in time [Fig.~\ref{fig:trajectories}(a,b)]. There are an infinite number of such steady states, corresponding to different initial conditions [Fig.~\ref{fig:trajectories}(c,d)]. We call this the oscillatory phase. (Note that these periodic orbits are \emph{not} limit cycles but are due to center fixed points \cite{strogatz01}.)

These periodic oscillations are reminiscent of those in Refs.~\cite{drummond78,kilin78}. However, a notable difference is that the frequency of the oscillations here depends on the initial condition [Fig.~\ref{fig:trajectories}(a,b)]. In general, a periodic orbit that covers more area on the Bloch sphere has a lower frequency.

To get some insight into the periodic oscillations, we note that according to Eqs.~\eqref{eq:coll_x} and \eqref{eq:coll_y}, there is a constant of motion,
\begin{eqnarray}
(X+Y)^{\left(\frac{2V}{\gamma_c}+1\right)}(X-Y)^{\left(\frac{2V}{\gamma_c}-1\right)}=C, \label{eq:const_motion}
\end{eqnarray}
where $C$ depends on the initial conditions. Thus, the trajectory on the Bloch sphere is given by the intersection of vertical sheets defined by Eq.~\eqref{eq:const_motion} with the sphere defined by $X^2+Y^2+Z^2=1$. Different initial conditions lead to different trajectories [Fig.~\ref{fig:trajectories}(c,d)]. There are four families of trajectories, corresponding to four quadrants in the $xy$-plane, since Eqs.~\eqref{eq:coll_x}--\eqref{eq:coll_z} have three symmetries: $(X,Y)\rightarrow (-X,-Y)$, $(X,Y)\rightarrow (Y,X)$, $(X,Y)\rightarrow (-Y,-X)$. Note that all trajectories come close to $X=Y=\pm \frac{1}{\sqrt{2}},Z=0$.

The nonsinuisoidal shape of the oscillations reflects the fact that the mean-field equations are nonlinear. The ``sawtooth'' shape of $Z(t)$ can be understood from Eq.~\eqref{eq:coll_z}. During the part of the cycle when $Z$ increases, the interaction tries to raise $Z$ while the decay tries to lower $Z$. During the part of the cycle when $Z$ decreases, both the interaction and the decay try to lower $Z$. Thus, $Z$ increases slower than it decreases.

When averaged over time, the oscillations satisfy
\begin{eqnarray}
\langle X\rangle_t\neq 0,\quad \langle Y\rangle_t\neq 0, \quad \langle Z\rangle_t=0, \label{eq:xyz_avg}
\end{eqnarray}
showing that the oscillatory phase breaks the $Z_2$ symmetry. The transition from PM to the oscillatory phase is first-order since the time-averaged values of $X,Y,Z$ are discontinuous there [Fig.~\ref{fig:jz}(b)].

Note that these results assume that the decay is purely collective. As mentioned in Sec.~\ref{sec:model}, in practice, there will always be a little bit of independent decay, which does not conserve angular momentum. Thus, for sufficiently long time scales, the oscillations will be modified by independent decay, and the amplitude may decrease over time.

In the limit of large $N$, one can again calculate the spin-squeezing parameter analytically by considering fluctuations around the mean-field steady states. In the PM, it is (see Appendix \ref{sec:app})
\begin{eqnarray}
\xi^2&=&\frac{\gamma_c}{\gamma_c+2|V|}, \quad\quad |V|<\frac{\gamma_c}{2}, \label{eq:xi_coll}
\end{eqnarray}
which is the same as for independent decay [Eq.~\eqref{eq:xi_indep}]. Thus, even with collective decay, the squeezing is limited to $\xi^2\geq 1/2$. In the oscillatory phase, there is no spin squeezing ($\xi^2>1$) because the periodic orbits are spread out across the Bloch sphere.


\subsection{Comparison with independent decay}

Comparing the results for independent decay and collective decay, we find that the phase boundaries are the same, $|V|=\gamma/2$. However, independent decay has a second-order transition while collective decay has a first-order transition. Also, while both models have the same PM phase, independent decay leads to a time-independent FM, while collective decay leads to a time-dependent oscillatory phase.

The fact that the phase boundaries are the same is due to the fact that in the PM, the Bloch vector points downward. In this regime,  collective decay has no collective enhancement \cite{gross82}, so it is effectively the same as independent decay. This also explains why spin squeezing is the same for both.

The intuition for the oscillatory phase is as follows. From Eq.~\eqref{eq:coll_z}, we see that the dissipation is maximum when $Z=0$, i.e., when the Bloch vector is on the equator. In order for the periodic orbits to exist, the Bloch vector must be able to pass above the equator: $\dot{Z}=2VXY-\gamma_c/2>0$. The first term is maximum when $X=Y=1/\sqrt{2}$, so to pass the equator, $V>\gamma_c/2$ must be satisfied, i.e., the interaction should be strong enough to lift the Bloch vector further. This is precisely when the oscillatory phase exists.

But why are there no periodic orbits for independent decay? Applying the same argument to Eq.~\eqref{eq:indep_z}, one requires $\dot{Z}=2VXY-\gamma_i>0$. The key difference is that with independent decay, the Bloch vector does not have maximal length, so $2VXY>\gamma_i$ is never satisfied. In fact, as $V$ increases, the Bloch vector becomes shorter due to decoherence, as seen from Eqs.~\eqref{eq:xss_indep}--\eqref{eq:zss_indep}. In other words, collective decay gives rise to periodic orbits because the conservation of angular momentum causes the Bloch vector to have maximal length. This also explains why collective decay has a first-order transition while independent decay has a second-order transition.

It is also interesting to compare our results with those of Refs.~\cite{morrison08a,morrison08b}, which studied a similar model with collective decay. Those papers found a second-order transition to a time-independent FM, in contrast to the first-order transition to an oscillatory phase here. However, those papers also found that $\xi^2$ reached a minimum value of 1/2 at the phase transition.

\subsection{Comparison with master equation} \label{sec:master}

Mean-field theory predicts an infinite number of steady states in the oscillatory phase. However, the original master equation [Eq.~\eqref{eq:master_coll}] has a unique steady-state density matrix $\rho_{ss}$. This discrepancy is resolved by noting that $\rho_{ss}$ is a mixture of all mean-field steady states. Thus, we expect $\rho_{ss}$ to satisfy $\langle J_x\rangle=\langle J_y\rangle=\langle J_z\rangle=0$ when $|V|>\gamma_c/2$. (Note that the Bloch vector calculated using $\rho_{ss}$ does not have maximal length in the oscillatory phase due to averaging over multiple mean-field steady states.)

To check the mean-field predictions, we numerically solve the master equation [Eq.~\eqref{eq:master_coll}] via Runge-Kutta integration to obtain $\rho_{ss}$. Figure \ref{fig:jz}(b) shows that when $N$ is large, there is a sudden change in $\langle J_z\rangle$ at $\Omega=\gamma_c/2$, which is consistent with a first-order transition there. As $N$ increases, it becomes more discontinuous. In general, mean-field theory is more accurate as $N$ increases, because fluctuations like $\langle J_x^2\rangle/N^2$ scale as $\sim 1/N$, as seen from Eqs.~\eqref{eq:Jxp2}--\eqref{eq:JxpJyp}.

\begin{figure}[tb]
\centering
\includegraphics[width=3 in,trim=0.in 0in 0in 0in,clip]{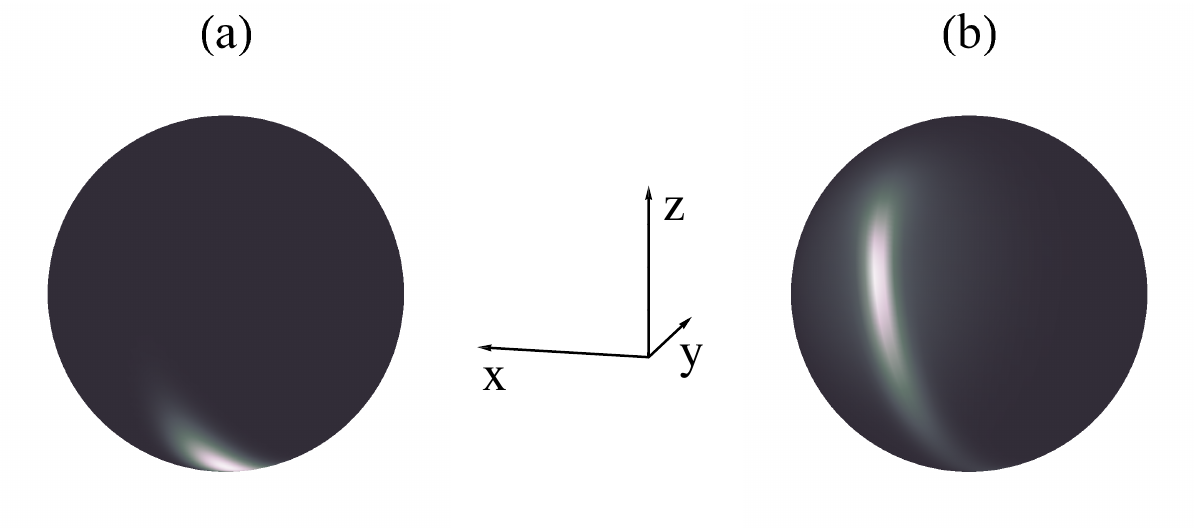}
\caption{\label{fig:wigner_V}Wigner function of the steady state for $N=50$ atoms. (a) $V=0.4\gamma_c$ and (b) $V=0.6\gamma_c$. Black denotes 0, while white denotes the maximum value. Due to the $Z_2$ symmetry, the Wigner function in panel (b) has another peak on the opposite side of the Bloch sphere. Compare panel (b) with Fig.~\ref{fig:trajectories}(c).}
\end{figure}

To visualize $\rho_{ss}$, we plot the Wigner function \cite{dowling94}. Figure \ref{fig:wigner_V}(a) shows what should be the PM phase; as expected, the Bloch vector points in the $-\hat{z}$ direction. Figure \ref{fig:wigner_V}(b) shows what should be the oscillatory phase; the Wigner function has peaks at $X=Y=\pm \frac{1}{\sqrt{2}},Z=0$, which is consistent with the fact that all the mean-field periodic orbits pass by those points [Fig.~\ref{fig:trajectories}(c)]. It seems that the Wigner function is always positive for large $N$.

\section{Addition of a drive} \label{sec:drive}

\begin{figure}[bt]
\centering
\includegraphics[width=3 in,trim=0.in 0in 0in 0in,clip]{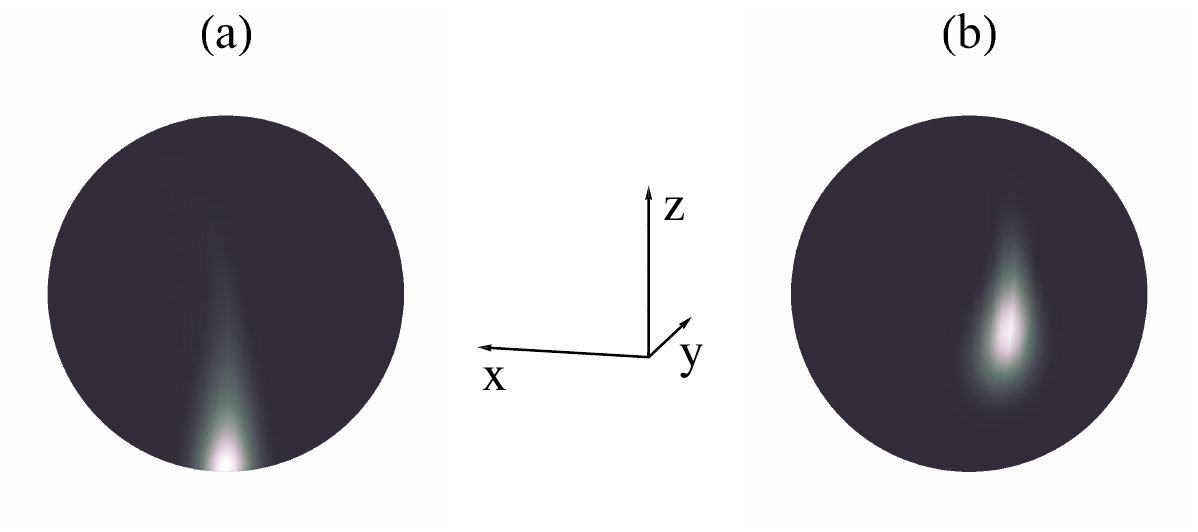}
\caption{\label{fig:wigner_VxW}Wigner function of the steady state for $N=50$ atoms. (a) $V_x=2\gamma_c$ and $\Omega=0$. (b) $V_x=0$ and $\Omega=0.45\gamma_c$. Black denotes 0, while white denotes the maximum value.}
\end{figure}

We would like to get more spin squeezing, but apparently the Hamiltonian Eq.~\eqref{eq:H_xy} with either independent or collective decay is not enough to decrease $\xi^2$ past 1/2. This is because the Bloch vector points downward in the PM and is unable to support much squeezing. To get around this, we add an external drive so that the Bloch vector points sideways.  We also omit the $J_y^2$ term, so that the Hamiltonian is
\begin{eqnarray}
H&=&\frac{V_x}{N}J_x^2 + \Omega J_x, \label{eq:H_drive}
\end{eqnarray}
and assume collective decay [Eq.~\eqref{eq:master_coll}]. The motivation for choosing this Hamiltonian is as follows. The first term $J_x^2$ is the one-axis twisting Hamiltonian \cite{kitagawa93}, which squeezes the spins in the $yz$ plane [Fig.~\ref{fig:wigner_VxW}(a)]. The second term causes the spins to precess around the $x$ axis; combined with the collective decay, this precession also leads to squeezing in the $yz$ plane [Fig.~\ref{fig:wigner_VxW}(b)]. Since both terms squeeze in the same plane, the hope is that the combination of the two leads to more squeezing than each by itself. This turns out to be true.

We note that this model with $V_x=0$ was previously studied in terms of its steady states \cite{drummond78,kilin78}, entanglement \cite{schneider02,morrison08c}, and spin squeezing \cite{gonzalez13,wolfe14b}. Here, we show analytically that $\xi^2\rightarrow0$ at the critical point and that the squeezing can be enhanced by adding interactions ($V_x\neq0$). We also find the scaling of $\xi^2$ with $N$.


\subsection{Mean-field equations}

The mean-field equations for this model are
\begin{eqnarray}
\dot{X}&=&\frac{\gamma_c}{2}XZ,\label{eq:drive_x}\\
\dot{Y}&=&-V_x XZ -\Omega Z + \frac{\gamma_c}{2}YZ,\\
\dot{Z}&=&V_x XY + \Omega Y- \frac{\gamma_c}{2}(1-Z^2).\label{eq:drive_z}
\end{eqnarray}
There is a phase transition when $\Omega=\gamma_c/2$. When $\Omega<\gamma_c/2$, the steady state is
\begin{eqnarray}
\bar{X}=0, \quad \bar{Y}=\frac{2\Omega}{\gamma_c}, \quad \bar{Z}=-\sqrt{1-\frac{4\Omega^2}{\gamma_c^2}}. \label{eq:ss_drive}
\end{eqnarray}
When $\Omega>\gamma_c/2$, the steady states are periodic orbits. We focus on the regime $\Omega<\gamma_c/2$, since that is where $\xi^2<1$.

Note that the steady state [Eq.~\eqref{eq:ss_drive}] and the critical point ($\Omega=\gamma_c/2$) are independent of $V_x$. This means we can make a clean comparison between $V_x=0$ and $V_x\neq 0$ in terms of spin squeezing.

\subsection{Spin squeezing}

By calculating fluctuations around the mean-field steady state [Eq.~\eqref{eq:ss_drive}], we find that in the limit of large $N$, the spin-squeezing parameter is (see Appendix \ref{sec:app})
\begin{eqnarray}
\xi^2=\frac{\gamma_c^2+V_x^2-2\Omega^2-2\sqrt{\gamma_c^2 V_x^2/4 + V_x^4/4+\Omega^4}}{\gamma_c\sqrt{\gamma_c^2-4\Omega^2}}, \quad\label{eq:squeezing_drive}
\end{eqnarray}
when $\Omega<\gamma_c/2$. We find that as $V_x$ increases, $\xi^2$ decreases monotonically.

\begin{figure}[tb]
\centering
\includegraphics[width=3. in,trim=1.in 3.3in 1in 3.4in,clip]{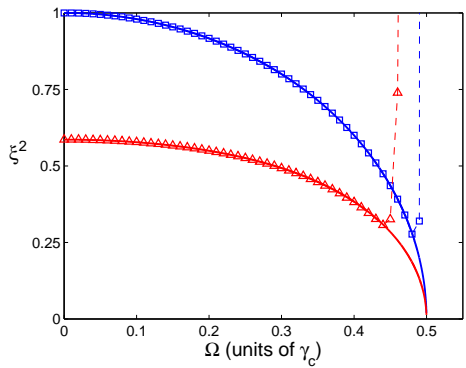}
\caption{\label{fig:squeezing_drive_W}Spin-squeezing parameter as a function of drive strength $\Omega$ for $V_x=0$ (blue squares) and $V_x=\gamma$ (red triangles). Solid lines are the analytical result Eq.~\eqref{eq:squeezing_drive}. Dashed lines and squares and triangles are the numerical results for $N=1000$, which deviate from the analytical predictions near the critical point due to finite-size effects.}
\end{figure}

When $V_x=0$,
\begin{eqnarray}
\xi^2&=&\sqrt{1-\frac{4\Omega^2}{\gamma_c^2}}. \label{eq:xi_Vx0}
\end{eqnarray}
When $V_x\rightarrow\infty$,
\begin{eqnarray}
\xi^2&=&\frac{1}{2}\sqrt{1-\frac{4\Omega^2}{\gamma_c^2}}.
\end{eqnarray}
At the critical point ($\Omega=\gamma_c/2$), the squeezing diverges ($\xi^2\rightarrow 0$), as seen in Fig.~\ref{fig:squeezing_drive_W}. Thus, adding a drive allows one to get a lot more squeezing. Furthermore, by setting $V_x$ large, one can get twice as much squeezing than with $V_x=0$.

Although squeezing diverges at the critical point, there are two points to be aware of. The first is that $\xi\rightarrow 0$ only when $N$ is infinite, since that is when mean-field theory is exact. When $N$ is finite, there will be finite-size effects. We discuss this further in Sec.~\ref{sec:finite}. The second point is that there will always be some independent decay in practice, which limits the squeezing. This is discussed further in Sec.~\ref{sec:limitations}.



\begin{figure}[tb]
\centering
\includegraphics[width=3.3 in,trim=1.1in 3.8in 1.3in 3.8in,clip]{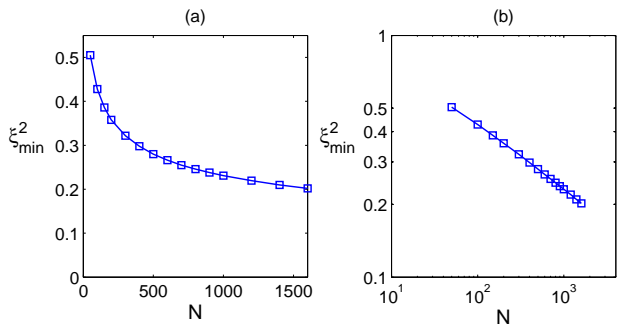}
\caption{\label{fig:squeezing_drive_N}Minimum spin-squeezing parameter as a function of $N$ for $V_x=0$. (Minimized with respect to $\Omega$.) (a) Linear scale. (b) Log-log scale.}
\end{figure}

\subsection{Limitations due to finite size} \label{sec:finite}

To study finite-size effects, we numerically integrate the master equation for $N=1000$ and calculate $\xi^2$. Figure \ref{fig:squeezing_drive_W} shows that there is good agreement with the analytical prediction, except near the critical point where there is an upturn in $\xi^2$. As $N$ increases, the agreement extends closer to the critical point. The reason for the discrepancy is that when $N$ is finite, mean-field theory is not self-consistent near the critical point since the fluctuations diverge there. [This can be seen by solving Eqs.~\eqref{eq:Jxp2}--\eqref{eq:JxpJyp}. See also Ref.~\cite{lee13d}.] However, since the fluctuations scale as $1/N$, mean-field theory becomes self-consistent again when $N$ is very large.

Figure \ref{fig:squeezing_drive_N} shows the minimum value of $\xi^2$ as a function of $N$. The data for $V_x=0$ suggest a power-law scaling,
\begin{eqnarray}
\xi^2_\text{min} &\approx& 1.70\,\, N^{-0.29}, \label{eq:xi2_N}
\end{eqnarray}
for large $N$. This scaling indicates that the squeezing here does not reach the Heisenberg limit ($\xi^2= 1/N$).

\subsection{Limitations due to independent decay} \label{sec:limitations}

To estimate the effect of independent decay, we first calculate the time scale required to reach the steady state of Eqs.~\eqref{eq:drive_x}--\eqref{eq:drive_z}. The time scale can be estimated within mean-field theory by linearizing Eqs.~\eqref{eq:drive_x}--\eqref{eq:drive_z} around the steady state [Eq.~\eqref{eq:ss_drive}] and calculating the stability eigenvalues. The relevant eigenvalue is
\begin{eqnarray}
\lambda&=&-\frac{\gamma_c}{2}\sqrt{1-\frac{4\Omega^2}{\gamma_c^2}}, \label{eq:lambda_Vx0}
\end{eqnarray}
so the time scale is
\begin{eqnarray}
\tau&=&\frac{2}{\gamma_c\sqrt{1-\frac{4\Omega^2}{\gamma_c^2}}}. \label{eq:time_scale}
\end{eqnarray}
The time scale diverges at the critical point ($\Omega=\gamma_c/2$), leading to a critical slowing down. In practice, one would not work exactly at the critical point, since it would take a very long time to reach the steady state. Also note that Eq.~\eqref{eq:time_scale} is independent of $V_x$; this means that by increasing $V_x$, one gets more squeezing \emph{without} having to wait longer.

Now we make a rough estimate of the effect of independent decay, similar to Ref.~\cite{hammerer04}. Suppose the independent decay rate $\gamma_i$ is very small: $\gamma_i\tau\ll 1$. During the time $\tau$ that it takes to reach the steady state of Eqs.~\eqref{eq:drive_x}--\eqref{eq:drive_z}, about $N\gamma_i\tau$ atoms have undergone an independent decay event, while $N(1-\gamma_i\tau)$ atoms have not. The former atoms have no squeezing ($\xi^2\approx 1$). The latter atoms have the squeezing given by Eq.~\eqref{eq:xi2_N}, which we call $\xi^2_0$. The overall squeezing of the atoms is estimated as a weighted average of the two groups:
\begin{eqnarray}
\xi^2_\text{total}&\approx& \xi^2_0 + \gamma_i\tau.
\end{eqnarray}
We write this in terms of the single-atom cooperativity, $\mathcal{C}=\frac{g^2}{\kappa\gamma_i}$, using the fact that $\gamma_c=N\mathcal{C}\gamma_i$ \cite{bohnet12}:
\begin{eqnarray}
\xi^2_\text{total}&\approx& \xi^2_0 + \frac{\gamma_c}{NC} \frac{1}{|\lambda|},\\
&\approx&\xi^2_0 + \frac{2}{NC\xi^2_0}, \label{eq:xi2_total}
\end{eqnarray}
since  Eqs.~\eqref{eq:xi_Vx0} and \eqref{eq:lambda_Vx0} show that $|\lambda|=\gamma_c\xi^2_0/2$. Since $\xi^2_0\sim N^{-0.29}$ according to Eq.~\eqref{eq:xi2_N}, for large $N$ or $C$, Eq.~\eqref{eq:xi2_total} is dominated by the first term.

As a concrete example, suppose there are $N=10^4$ atoms in a cavity with $\mathcal{C}=0.1$. For simplicity, let $V_x=0$. We use a level scheme like in Ref.~\cite{bohnet12} so that $\gamma_i$ can be set to a small value. We assume $\gamma_i=2\pi\times 10\text{ kHz}$ so that $\gamma_c=2\pi\times 10\text{ MHz}$. From Eqs.~\eqref{eq:xi2_N} and \eqref{eq:xi2_total}, we find $\xi^2_0=0.12$ and $\xi^2_\text{total}=0.13$. The time to reach steady state is about $\tau=0.3\; \mu\text{s}$. The critical point occurs at $\Omega=\gamma_c/2=2\pi\times 5\text{ MHz}$. Due to the steep slope near the critical point in Fig.~\ref{fig:squeezing_drive_W}, $\Omega$ needs to be relatively precise.

\section{Conclusion}

We have shown that collective and independent decay lead to qualitatively different phase transitions. The differences are ultimately due to the fact that collective decay conserves angular momentum, which causes the Bloch vector to have maximal length and enables the oscillatory phase to exist. Then we showed that adding a drive allows for infinite spin squeezing because the Bloch vector points sideways. For future work, it would be interesting to consider what happens when the spin-spin interaction is short range but the decay is still collective. One can also consider the effect of a non-Markovian environment \cite{chan14}. Another promising direction is to generate spin squeezing using a non-Hermitian Hamiltonian instead of a master equation \cite{lee14b}.




\section{Acknowledgements}
We thank Monika Schleier-Smith and Florentin Reiter for useful comments. This work was supported by NSF.

\appendix

\section{Calculation of spin-squeezing parameter} \label{sec:app}

In this appendix, we calculate the spin-squeezing parameter in the limit of large $N$ by considering fluctuations around the mean-field steady state using the approach of Ref.~\cite{morrison08b}. For generality, we use the Hamiltonian
\begin{eqnarray}
H&=&\frac{V_x}{N}J_x^2 + \frac{V_y}{N}J_y^2 + \Omega J_x,\label{eq:H_app}
\end{eqnarray}
which encompasses Eqs.~\eqref{eq:H_xy} and \eqref{eq:H_drive}. We assume collective decay [Eq.~\eqref{eq:master_coll}].

The mean-field equations are:
\begin{eqnarray}
\dot{X}&=&V_y YZ + \frac{\gamma_c}{2}XZ,\\
\dot{Y}&=&-V_x XZ - \Omega Z + \frac{\gamma_c}{2}YZ,\\
\dot{Z}&=&(V_x-V_y)XY +\Omega Y - \frac{\gamma_c}{2}(1-Z^2).
\end{eqnarray}
A phase transition occurs when $\Omega=\Omega_c$, where
\begin{eqnarray}
\Omega_c\equiv\frac{\gamma_c^2 + 4 V_x V_y}{2\sqrt{\gamma_c^2 + 4 V_y^2}}.
\end{eqnarray}
When $\Omega<\Omega_c$, the steady state is the fixed point,
\begin{eqnarray}
\bar{X}&=&-\frac{4 V_y \Omega}{\gamma_c^2 + 4 V_x V_y} = \sin\theta \cos\phi, \label{eq:ss_x_app}\\
\bar{Y}&=& \frac{2\gamma_c\Omega}{\gamma_c^2 + 4 V_x V_y} = \sin\theta \sin\phi, \\
\bar{Z}&=& -\frac{\sqrt{(\gamma_c^2 + 4 V_xV_y)^2 - 4\Omega^2(\gamma_c^2 + 4 V_y^2)}}{\gamma_c^2 + 4 V_xV_y} = \cos\theta, \quad\quad\label{eq:ss_z_app}
\end{eqnarray}
where $\theta,\phi$ are the spherical angles of the Bloch vector. When $\Omega>\Omega_c$, there is no stable fixed point, and the steady states are periodic orbits. Below, we focus on $\Omega<\Omega_c$, since that is where spin squeezing exists.

To calculate the spin-squeezing parameter, it is convenient to first rotate the spin operators $\vec{J}$ by the angle $\theta$ around the axis $\hat{n}=(-\sin\phi, \cos\phi,0)$,
\begin{eqnarray}
\vec{J}'&=&e^{-i \theta \hat{n}\cdot \vec{J}} \vec{J} e^{i \theta \hat{n}\cdot \vec{J}}, \label{eq:Jprime}\\
e^{-i \theta \hat{n}\cdot \vec{J}} &=& \cos \frac{\theta}{2} \,I - 2i\sin\frac{\theta}{2}(-\sin\phi \,J_x + \cos\phi \,J_y).\quad\quad
\end{eqnarray}
The new spin operators $\vec{J}'$ are defined such that $J_z'$ points in the same direction as the Bloch vector of the steady state:
\begin{eqnarray}
\langle J_x'\rangle=\langle J_y'\rangle=0, \quad \langle J_z'\rangle=\frac{N}{2}.
\end{eqnarray}
In terms of $\vec{J'}$, the spin-squeezing parameter is \cite{wineland92}
\begin{eqnarray}
\xi^2&=&\frac{\langle J_x'^2 + J_y'^2\rangle - \sqrt{\langle J_x'^2 - J_y'^2\rangle^2 + \langle J_x'J_y' + J_y'J_x'\rangle^2}}{\frac{N}{2}}. \nonumber \\ \label{eq:xi2}
\end{eqnarray}
Since the Bloch vector has maximal length, the definitions of $\xi^2$ in Refs.~\cite{wineland92,kitagawa93} are identical.

So to calculate squeezing, we need to calculate fluctuations of $\vec{J}'$. This is most conveniently done by using the Holstein-Primakoff transformation to write $\vec{J'}$ in terms of bosonic annihilation and creation operators ($a,a^\dagger$). In the limit of large $N$,
\begin{eqnarray}
J_+' = \sqrt{N} a, \quad J_-' = \sqrt{N} a^\dagger, \quad J_z' = \frac{N}{2} - a^\dagger a, \quad\label{eq:hp}
\end{eqnarray}
where $[a,a^\dagger]=1$. Now we rewrite the master equation [Eq.~\eqref{eq:master_coll}] in terms of $a,a^\dagger$. First, we invert Eq.~\eqref{eq:Jprime} to find:
\begin{eqnarray}
J_{\pm}&=&\cos^2\frac{\theta}{2} J_\pm' - e^{\pm 2i\phi}\sin^2\frac{\theta}{2} J_\mp' + e^{\pm i\phi}\sin\theta J_z'. \quad\quad \label{eq:Jpm}
\end{eqnarray}
We substitute Eqs.~\eqref{eq:hp} and \eqref{eq:Jpm} into Eqs.~\eqref{eq:master_coll} and \eqref{eq:H_app}. The terms linear in $a,a^\dagger$ cancel out due to the definition of $\vec{J'}$, so the leading order is quadratic. We keep only quadratic terms, since we are interested in the limit of large $N$. The resulting master equation is
\begin{eqnarray}
\dot{\rho}&=&-i[H,\rho]+\frac{\gamma_c}{2} \bigg[\cos^4\frac{\theta}{2}(2a^\dagger\rho a - aa^\dagger\rho - \rho aa^\dagger) \nonumber\\
&& \quad\quad\quad\quad\quad\quad + \sin^4\frac{\theta}{2}(2a\rho a^\dagger - a^\dagger a\rho - \rho a^\dagger a) \nonumber\\
&& \quad\quad\quad\quad\quad\quad - \frac{1}{4}e^{-2i\phi}\sin^2\theta(2a\rho a - a^2\rho - \rho a^2) \nonumber\\
&& \quad\quad\quad\quad\quad\quad- \frac{1}{4}e^{2i\phi}\sin^2\theta(2a^\dagger\rho a^\dagger - a^{\dagger2}\rho - \rho a^{\dagger2})\bigg], \nonumber\\ \\
H&=& b_1 a^2 + b_1^* a^{\dagger2} +  b_2 a^\dagger a, \\
b_1&=& \frac{e^{-2i\phi}}{4}[V_x (\cos\theta\cos\phi + i\sin\phi)^2 \nonumber\\
&& \quad\quad\quad- V_y (\cos\phi + i\cos\theta\sin\phi)^2], \\
b_2&=& \frac{1}{8}[V_x + V_y + 3(V_x+V_y)\cos(2\theta) - 8\Omega \sin\theta\cos\phi \nonumber\\
&& \quad + 6(-V_x+V_y)\sin^2\theta\cos(2\phi)].
\end{eqnarray}
(We note that a similar master equation occurs in the context of a cavity mode coupled to a quantum dot \cite{zhu14}.)
Since there are no terms linear in $a,a^\dagger$, we have $\langle a\rangle=\langle a^\dagger\rangle=0$. The equations of motion for the fluctuations are
\begin{eqnarray}
\frac{d\langle a^2\rangle}{dt}&=&-2ib_1^*(2\langle a^\dagger a\rangle + 1)-2ib_2\langle a^2\rangle \nonumber\\
&& \quad +\gamma_c\left(\cos^4\frac{\theta}{2}-\sin^4\frac{\theta}{2}\right)\langle a^2\rangle + \frac{\gamma_c}{4}e^{2i\phi}\sin^2\theta, \nonumber\\ \\
\frac{d\langle a^{\dagger2}\rangle}{dt}&=&2ib_1(2\langle a^\dagger a\rangle + 1)+2ib_2\langle a^{\dagger2}\rangle \nonumber\\
&& \quad +\gamma_c\left(\cos^4\frac{\theta}{2}-\sin^4\frac{\theta}{2}\right)\langle a^{\dagger2}\rangle + \frac{\gamma_c}{4}e^{-2i\phi}\sin^2\theta, \nonumber\\ \\
\frac{d\langle a^\dagger a\rangle}{dt}&=& 2i(b_1 \langle a^2\rangle - b_1^* \langle a^{\dagger2}\rangle) \nonumber\\
&& \quad +\gamma_c\left(\cos^4\frac{\theta}{2} - \sin^4\frac{\theta}{2}\right)\langle a^\dagger a\rangle + \gamma_c \cos^4\frac{\theta}{2}. \nonumber \\
\end{eqnarray}
We solve these equations for the steady-state fluctuations. The resulting expressions are complicated, so we do not write them out here.

Then we convert fluctuations of $a,a^\dagger$ into fluctuations of $\vec{J}'$:
\begin{eqnarray}
\langle J_x'^2\rangle&=&\frac{N}{4}(\langle a^2\rangle + \langle a^{\dagger2}\rangle + 2\langle a^\dagger a\rangle + 1), \quad\quad\label{eq:Jxp2}\\
\langle J_y'^2\rangle&=&-\frac{N}{4}(\langle a^2\rangle + \langle a^{\dagger2}\rangle - 2\langle a^\dagger a\rangle - 1), \quad\quad\label{eq:Jyp2}\\
\langle J_x'J_y' + J_y'J_x'\rangle&=&-\frac{iN}{2}(\langle a^2\rangle - \langle a^{\dagger2}\rangle).\label{eq:JxpJyp}
\end{eqnarray}
We plug these expressions into Eq.~\eqref{eq:xi2} to find $\xi^2$. The resulting expression for $\xi^2$ is very complicated. However, for the special cases considered in the main text, we get relatively simple expressions [Eqs.~\eqref{eq:xi_coll} and \eqref{eq:squeezing_drive}].

\bibliography{collective}

\end{document}